\documentclass[%
 aip,
 jmp,%
 amsmath,amssymb,
preprint,%
]{revtex4-1}

\usepackage{graphicx}
\usepackage{subfigure}
\usepackage[titletoc,toc,title,page]{appendix}
\usepackage{placeins}
\usepackage{color}
\usepackage{dcolumn}
\usepackage{bm}

\begin{document}

\title{The interaction of seasonality and low-frequencies in a stochastic Arctic sea ice model}

\author{
W. Moon} 

\affiliation{British Antarctic Survey, High Cross, Madingley Rd, Cambridge CB3 0ET, United Kingdom
}%
 \email{woosok.moon@gmail.com}

\begin{abstract}
The stochastic Arctic sea ice model described as a single periodic non-autonomous stochastic ordinary differential equation (ODE) is useful in explaining the seasonal variability of Arctic sea ice. 
However, to be nearer to realistic approximations we consider the inclusion 
of long-term forcing implying the effect of slowly-varying ocean or atmospheric low-frequencies. 
In this research, we rely on the equivalent Fokker-Planck equation instead of the stochastic ODE owing to the advantages of the Fokker-Planck equation 
 in dealing with higher moments calculations. We include simple long-term forcing into the Fokker-Planck equation and then 
seek approximate stochastic solutions. The formalism based on the Fokker-Planck equation with a singular perturbation method is flexible with regard to accommodating further 
complexity that arises due to the inclusion of long-term forcing. These solutions are then applied to
the stochastic Arctic sea ice model with long-term forcing. Strong seasonality in the Arctic sea ice model combined with 
long-term forcing, changes the seasonal variability depending on the phase of the long-term forcing. The change includes the shift of mean and
the increase or decrease of variance and skewness. Stochastic realisations show that the change of the statistical moments due to long-term forcing 
is realised by unusual fluctuations particularly concentrated at a specific time of a year.  
\end{abstract}

%

\maketitle

\section{Introduction}  
A periodic non-autonomous stochastic ordinary differential equation (ODE) can be used to study seasonal variability in the field of climate science. 
In particular, the study of the seasonal variability of Arctic sea ice, which shows large seasonal fluctuations, involves considering
a periodic non-autonomous stochastic ODE as a first order differential equation \cite{EW09, MW:2011, MW:2012}. However, the climate also contains slowly varying variabilities 
caused by ocean circulations or large-scale atmospheric 
low-frequencies \cite{Branstator1992, Hurrell1995}, owing to which 
the inclusion of a long-term forcing should be considered into the related periodic
non-autonomous stochastic ODE to be more near to actual climate variability.
Along these lines, one of the key issues is determining how long-term forcing interacts with seasonality, and accordingly,
influences seasonal variability. 

Stochastic solutions without long-term forcing have been constructed in previous studies using a regular perturbation method based on the assumption
that noise magnitude is much lesser than the degree of seasonal cycle \cite{MW:2013}. 
These approximate solutions reveal several important physical characteristics of 
seasonal variability, which is explained by the interaction between seasonal stability and noise forcing. In particular, the accumulation 
effect of the responses to noise forcing controlled by seasonal stability is the main physics towards understanding the seasonal evolution of stochastic solutions.  

The best example for the above generalised perspective is the seasonal variability of Arctic sea ice. Beginning from early summer, Arctic sea ice becomes
less stable or unstable due to sea ice albedo feedback, which implies that a decline in sea ice albedo induced by melting sea ice leads to more melting.  
During mid-summer, short-wave radiance increases rapidly in the Arctic basin, and then, the intensity of the sea-ice albedo feedback is maximised. 
One might expect that the 
variance of Arctic sea ice thickness is maximised in the middle of summer considering the stability and the noise magnitude; however, the variance of sea ice thickness is maximised nearer to the end of summer. The responses of Arctic sea ice to
noise forcing accumulates until the sea ice albedo feedback almost vanishes, which normally happens nearer to the end of summer. 
This could be termed as the "memory effect",
which is well described in a previous research based on a delayed-integral form of stochastic solutions \cite{MW:2013}.

In spite of the success of the previous research in capturing the core of the seasonal variability of Arctic sea ice, 
we have to consider a more realistic situation, i.e., the seasonal cycle is influenced by various long-term forcing including ocean circulations or large-scale atmospheric 
low-frequencies. Multi-fractal time-series analysis reveals that decadal time scales exist separately after showing the white noise 
characteristics between seasonal and bi-seasonal in Arctic sea ice extent data \cite{Agarwal 2012}, which may not be a special characteristic of Arctic sea ice but 
a more general one in any climate phenomenon. 
Hence, we can consider a minimal comprehensive model to address the issue regarding the interaction of low-frequencies and seasonal stochasticity.

One possible approach is to consider simple long-term forcing represented as sinusoidal functions
in the seasonal stochastic model. Mathematically, it is the inclusion of an additional forcing term, 
whose magnitude slowly increases with time.
Seasonal variability could be different depending on the phase of the periodic low frequency. 
To analyse this effect explicitly, we need to find approximate solutions that include long-term forcing and then compare them with the original stochastic model without long-term forcing. 

Based on the views discussed above, this research focusses on the construction of stochastic solutions 
for a periodic non-autonomous stochastic model that includes low-frequency forcing. Previously, a regular perturbation method was applied to a given stochastic ODE based on stochastic calculus \cite{MW:2013}. 
The shortcoming of the regular perturbation method for a stochastic ODE is that one would need to deal with complicated integral equations of higher orders. 
To overcome this issue in the regular perturbation method,  
we use the equivalent Fokker-Planck equation, which provides us with a more direct and convenient method to calculate higher moments. 

Physically, the main issue is how long-term forcing interacts with seasonality and noise forcing. Depending on the phase 
of the long-term forcing, the interaction could cause an increase or decrease in seasonal variability. This research investigates the influence the interaction with long-term forcing has on seasonal
variability 
in detail, with regard to the stochastic model. Moreover, we focus on the possibility that an increase in the seasonal variability is related to the occurrence of 
extreme events during a specific time period in a year. 

The remainder of this paper is organised as follows, in section \ref{sec:model}, the formalism based on the Fokker-Planck equation is introduced. We show that this equivalent method also 
provides the same approximate solutions as that in previous research. The inclusion of long-term forcing is considered
using the Fokker-Planck equation in section \ref{sec:longterm}. Here, we systematically interpret the new terms generated by the inclusion of long-term forcing. 
The solutions and the general interpretations of the given stochastic solutions are applied to the Arctic sea ice model in section \ref{sec:seaice}. Finally, the implications 
of the results are discussed in the conclusion.

\section{\label{sec:model} Stochastic perturbation theory based on Fokker-Planck equation}
A periodic non-autonomous 1-dimensional stochastic differential equation is generally represented as 
\begin{align} \label{eqn:sto01}
 \frac{dX}{dt} = a(X,t)+\sigma b(X,t) \xi,
\end{align}
where $\xi$ is white noise and $a(X,t)$ and $b(X,t)$ are periodic functions.
Under the assumption that we already know the steady-state solution, $X_S$, of the deterministic part and also 
that the magnitude of the noise term, $\sigma$, is small,
we can consider a perturbation, $\eta$, around $X_S$ 
such that we can put $X=X_S+\eta$ into Eq. (\ref{eqn:sto01}), which leads to
\begin{align} \label{eqn:sto02}
 \frac{d\eta}{dt}=c(t)\eta+d(t)\eta^2+\sigma (N(t)+g(t)\eta)\xi
\end{align}
with
\begin{align} \label{eqn:term01}
c(t) &=\frac{\partial a}{\partial X}\arrowvert_{X=X_S}, \\ 
d(t) &=\frac{1}{2}\frac{\partial^2 a}{\partial X^2} \arrowvert_{X=X_S}, \\
N(t) &= b(X_S,t), \qquad \\
g(t) &= \frac{\partial b}{\partial X}\arrowvert_{X=X_S}.
\end{align}
The equivalent Fokker-Planck equation is
\begin{align} \label{eqn:fokker01}
 \frac{\partial P}{\partial t}=-\frac{\partial}{\partial \eta}\left(a(t)\eta+b(t)\eta^2\right)P+\frac{1}{2}\sigma^2\frac{\partial^2}{\partial\eta^2}\left(N(t)+g(t)\eta\right)^2 P.
\end{align}
We can rescale $\eta$ as $\eta=\sigma x$ and thereby obtain
\begin{align} \label{eqn:fokker02}
 \frac{\partial P}{\partial t} = &-\frac{\partial}{\partial x}\left(a(t)x+\sigma b(t)x^2+\sigma N(t)g(t)\right)P \nonumber \\
                                         &+\frac{1}{2}\frac{\partial^2}{\partial x^2}\left(N^2(t)+2\sigma N(t)g(t)+\sigma^2 g^2(t)x^2\right)P.
\end{align}
Considering a perturbation series using $\sigma$, $P=P_0+\sigma P_1 + \sigma^2 P_2 + \cdots$,
we obtain the following first two leading order equations,
\begin{align} \label{eqn:per_fokker01}
 &O(1) : \frac{\partial P_0}{\partial t} = -\frac{\partial}{\partial x}\left(a(t)x\right)P_0+\frac{1}{2}N^2(t)\frac{\partial^2 P_0}{\partial x^2} \\
 &O(\sigma) : \frac{\partial P_1}{\partial t} = -\frac{\partial}{\partial x}\left(a(t)x\right)P_1+\frac{1}{2}N^2(t)\frac{\partial^2 P_1}{\partial x^2}  \nonumber \\
                                                                  & -b(t)\frac{\partial}{\partial x}\left(x^2 P_0\right)-N(t)g(t)\frac{\partial P_0}{\partial x}
                                                                    +g(t)N(t)\frac{\partial^2}{\partial x^2}xP_0.
\end{align}
$O(1)$ is linear and the domain of $x$ is from $-\infty$ to $+\infty$, such that Fourier transform can be used,
\begin{align} \label{eqn:fourier}
 \hat{P_0} = \frac{1}{\sqrt{2\pi}}\int_{-\infty}^{+\infty}P_0e^{-ikx}dx.
\end{align}
The Fourier transformation for $O(1)$ is
\begin{align} \label{eqn:fourier_eqn01}
\frac{\partial \hat{P_0}}{\partial t} = a(t)k\frac{\partial \hat{P_0}}{\partial k} - \frac{1}{2}k^2 N^2(t) \hat{P_0}.
\end{align}
The characteristics method is applied to solve the above equation. 
The characteristic equations are 
\begin{align} \label{eqn:characteristic}
 &\frac{dk}{dt} = -a(t)k  \\
 &\frac{d\hat{P_0}}{dt}=-\frac{1}{2}k^2 N^2(t) \hat{P_0},
\end{align}
which leads to
\begin{align} \label{eqn:sol_character}
 \hat{P_0}(k,t)=C \text{exp}\left(-\frac{1}{2}\sigma^2_t k^2\right),
\end{align}
where $C$ is a constant and 
\begin{align}
 \sigma^2_t=\text{exp}\left(2\int_{0}^{t}a(r)dr\right)\int_{0}^{t} N^2(r)\text{exp}\left(-2\int_{0}^{r}a(s)ds\right)dr.
\end{align}
After taking the inverse Fourier transformation, the resulting solution is
\begin{align}
 P_0(x,t) = \frac{C}{\sqrt{2\pi\sigma^2_t}} \text{exp}\left(-\frac{x^2}{2\sigma^2_t}\right).
\end{align}

The Fourier transform is applied again to $O(\sigma)$, which leads to
\begin{align}
 \frac{\partial \hat{P_1}}{\partial t} &= - a(t)k\frac{\partial \hat{P_1}}{\partial k} - \frac{1}{2}k^2N^2(t)\hat{P_1} \nonumber \\
                                                     &+ikb(t)\frac{\partial^2 \hat{P_0}}{\partial k^2}-ikN(t)g(t)\hat{P_0}-ik^2N(t)g(t)\frac{\partial \hat{P_0}}{\partial k}.
\end{align}  
The resultant characteristic equations are
\begin{align}
 &\frac{dk}{dt} = -a(t)k \label{eqn:k} \\
 &\frac{d\hat{P_1}}{dt} = -\frac{1}{2}k^2N^2(t)\hat{P_1}+ib(t)k\frac{\partial^2\hat{P_0}}{\partial k^2}-iN(t)g(t)k\hat{P_0}-ig(t)N(t)k^2\frac{\partial \hat{P_0}}{\partial k}. \label{eqn:p1}
\end{align}
The solution in the Fourier domain is
\begin{align} \label{eqn:p1k}
 \hat{P_1}(k,t)=\frac{i}{\sqrt{2\pi}}\frac{1}{2}k^3(s_1+s_2)\text{exp}\left(-\frac{1}{2}k^2\sigma^2_t\right)
  -\frac{i}{\sqrt{2\pi}}k(m_1+m_2)\text{exp}\left(-\frac{1}{2}k^2\sigma^2_t\right),
\end{align}
where
\begin{align}
  &m_1 \equiv \text{exp}\left(\int_{0}^{t}a(r)dr\right)\int_{0}^{t}b(t')
          \text{exp}\left(\int_{0}^{t'}a(r)dr\right)\int_{0}^{t'}N^2(r)\text{exp}\left(-2\int_{0}^{r}a(s)ds\right)drdt' \nonumber \\
  &m_2 \equiv \text{exp}\left(\int_{0}^{t}a(r)dr\right)\int_{0}^{t}N(t')g(t')\text{exp}\left(-\int_{0}^{t'}a(r)dr\right)dt' \nonumber \\
  &s_1 \equiv 2\text{exp}\left(3\int_{0}^{t}a(r)dr\right)\int_{0}^{t}b(t')\text{exp}\left(\int_{0}^{t'}a(r)dr\right)
           \left(\int_{0}^{t'}N^2(r)\text{exp}\left(-2\int_{0}^{r}a(s)ds\right)dr\right)^2dt' \nonumber \\
  &s_2 \equiv 2\text{exp}\left(3\int_{0}^{t}a(r)dr\right)\int_{0}^{t}g(t')N(t')\text{exp}\left(-\int_{0}^{t'}a(r)dr\right)
                \int_{0}^{t'}N^2(r)\text{exp}\left(-2\int_{0}^{r}a(s)ds\right)dr dt'.
\end{align}
After taking the inverse Fourier transform of equation (\ref{eqn:p1k}), using the following relationships :
\begin{align}
\frac{1}{\sqrt{2\pi}}\int_{-\infty}^{+\infty}k^3 \text{exp}\left(-\frac{1}{2}k^2\sigma^2_t\right) e^{ikx} dk = i \frac{\partial^3 P_0}{\partial x^3} \nonumber \\
\frac{1}{\sqrt{2\pi}}\int_{-\infty}^{+\infty} k \text{exp}\left(-\frac{1}{2}k^2\sigma^2_t\right) e^{ikx} dk = - i \frac{\partial P_0}{\partial x},
\end{align}
we finally have
\begin{align}
 P_1(x,t) = -\frac{1}{2} (s_1+s_2) \frac{\partial^3 P_0}{\partial x^3} - (m_1+m_2)  \frac{\partial P_0}{\partial x}.
\end{align}
Hence,
\begin{align}
 P(x,t) = P_0(x,t) - \sigma \left(\frac{1}{2}(s_1+s_2)\frac{\partial^3 P_0}{\partial x^3}+(m_1+m_2)\frac{\partial P_0}{\partial x}\right).
\end{align}
Instead of $x$, $\eta$ can be used to express $P$,
\begin{align}
 P^{*}(\eta,t) = \frac{1}{\sigma}P=P^{*}_0-\sigma^2(m_1+m_2)\frac{\partial P^{*}_0}{\partial \eta}-\frac{1}{2}\sigma^4(s_1+s_2)\frac{\partial^3 P^{*}_0}{\partial \eta^3},
\end{align}
where
\begin{align}
 P^{*}_0 = \frac{1}{\sqrt{2\pi\sigma^2\sigma^2_t}}\text{exp}\left(-\frac{\eta^2}{2\sigma\sigma^2_t}\right).
\end{align}
To compare the results with that of the previous research \cite{MW:2013}, we calculate several statistical moments, which are
\begin{align}
&<\eta^2> = \int_{-\infty}^{+\infty}\eta^2 P^{*}_0 d\eta = \sigma^2\sigma^2_t \\
&<\eta> = -\sigma^2(m_1+m_2)\int_{-\infty}^{+\infty}\eta\frac{\partial P^{*}_0}{\partial \eta}d\eta = \sigma^2(m_1+m_2) \\
&<\eta^3> = 3\sigma^4(s_1+s_2) + 3\sigma^4(m_1+m_2)\sigma^2_t.
\end{align}
Therefore, we obtain $<(\eta-<\eta>)^3> = 3\sigma^4(s_1+s_2)+O(\sigma^6)$. 
These results are consistent with the previous research \cite{MW:2013}, which is based on a regular perturbation method in the original 
stochastic ODE.

\section{\label{sec:longterm} Inclusion of long-term forcing}
Now, we consider the inclusion of long-term forcing in the above stochastic model. In general, long-term forcing in a climate system 
can be a result of low-order chaotic systems.
However, we consider the simplest form of long-term forcing, periodic low-frequency forcing, which can form the basis to understanding more complicated cases. 

\subsection{\label{sec:periodic} Perturbation Method with simple periodic forcing}

The relevant Fokker-Planck equation with simple periodic forcing is
\begin{align}
\frac{\partial P}{\partial t}=-\frac{\partial}{\partial \eta}\left(a(t)\eta+b(t)\eta^2+c_1\sigma\text{cos}(\omega t)\right)P
                                       +\frac{1}{2}\sigma^2\frac{\partial^2}{\partial \eta^2}\left(N(t)+g(t)\eta\right)^2 P,
\end{align}
where $c_1\sigma\text{cos}(\omega t)$ is the simple periodic forcing satisfying 
$\omega << 1$ and $O(c_1) \sim O(1)$. It is important, here, to point out that periodic forcing has the same order of magnitude as noise forcing. 
Although we can choose any scale for the magnitude of the periodic forcing, 
it is more realistic to choose the same order as that of the starting point. The difficulty in detecting low-frequencies 
in data comes from the speculation that seasonal variability masks low-frequencies, which simply leads to the assumption that the magnitude of low-frequencies
is not bigger than that of seasonal variability. 

Rescaling $\eta$ as $\eta=\sigma x$ leads to
\begin{align}
 \frac{\partial P}{\partial t} &=-\frac{\partial}{\partial x}\left(a(t)x+\sigma b(t) x^2 + \sigma N(t)g(t) + c_1\text{cos}(\omega t)\right) P \nonumber \\
                   &+\frac{1}{2} \frac{\partial^2}{\partial x^2}\left(N^2(t)+2\sigma N(t)g(t)x+\sigma^2 g^2(t)x^2\right)P.
\end{align}
We can let $P=P_0+\sigma P_1+\sigma^2 P_2 + \cdots$ such that $O(1)$ becomes
\begin{align}
 \frac{\partial P_0}{\partial t}= -a(t)\frac{\partial}{\partial x}xP_0-c_1\text{cos}(\omega t)\frac{\partial P_0}{\partial x}+\frac{1}{2}N^2(t)\frac{\partial^2 P_0}{\partial x^2}.
\end{align}
In the frequency domain, equation 3.3 is transformed to
\begin{align}
 \frac{\partial \hat{P_0}}{\partial t}=a(t)k\frac{\partial \hat{P_0}}{\partial k}-ic_1k\text{cos}(\omega t)\hat{P_0}-\frac{1}{2}k^2N^2(t)\hat{P_0}.
\end{align}
The characteristic equations are
\begin{align}
 &\frac{dk}{dt} = -a(t)k \nonumber \\
 &\frac{d\hat{P_0}}{dt}=-\left(i c_1 k\text{cos}(\omega t)+\frac{1}{2}k^2 N^2(t)\right)\hat{P_0}.
\end{align}
We obtain
\begin{align}
 \hat{P_0}=C\text{exp}\left(-ikL-\frac{1}{2}\sigma^2_t k^2\right),
\end{align}
where
\begin{align}
& L \equiv c_1\text{exp}\left(\int_{0}^{t}a(r)dr\right)\int_{0}^{t}\text{cos}(\omega t')\text{exp}\left(-\int_{0}^{t'}a(r)dr\right)dt' \\
&\sigma^2_t \equiv \text{exp}\left(2\int_{0}^{t}a(r)dr\right)\int_{0}^{t}N^2(r)\text{exp}\left(-2\int_{0}^{r}a(s)ds\right)dr.
\end{align}
After talking the inverse Fourier transform, the solution obtained is as follows
\begin{align}
 P_0 = \frac{C}{\sigma_t}\text{exp}\left(-\frac{(x-L)^2}{2\sigma^2_t}\right),
\end{align}
where $C$ is a constant that is used for normalisation. 

The equation for the next order $O(\sigma)$ is 
\begin{align}
&\frac{\partial P_1}{\partial t}+\frac{\partial}{\partial x}\left(a(t)x+c_1\text{cos}(\omega t)\right)P_1 - \frac{1}{2}N^2(t)\frac{\partial^2 P_1}{\partial x^2} \nonumber \\
&=-b(t)\frac{\partial}{\partial x}\left(x^2+N(t)g(t)\right)P_0 + N(t)g(t)\frac{\partial^2}{\partial x^2}xP_0.
\end{align}
Using the characteristic equations, we can construct
\begin{align}
\hat{P_1}(k,t)&=iC\left[\frac{1}{2}k^3(s_1+s_2)-k(m_1+m_2+c^2_1m_3)+k^2(2ic_1 St_1+ic_1St_2)\right]
                       \text{exp}\left(-iLk-\frac{1}{2}\sigma^2_t k^2\right),
\end{align}
where
\begin{align}
St_1 = &\text{exp}\left(2\int_{0}^{t}a(r)dr\right)\int_{0}^{t}b(t')\text{exp}\left(\int_{0}^{t'}a(r)dr\right)\int_{0}^{t'}\text{cos}(\omega r)
            \text{exp}\left(-\int_{0}^{r}a(s)ds\right)dr \nonumber \\ 
            &\times \int_{0}^{t'} N^2(r)\text{exp}\left(-2\int_{0}^{r}a(s)ds\right)drdt'\\
St_2 = &\text{exp}\left(2\int_{0}^{t}a(r)dr\right)\int_{0}^{t}N(t')g(t')\text{exp}\left(-\int_{0}^{t'}a(r)dr\right) \nonumber \\
            &\times \int_{0}^{t'}\text{cos}(\omega r)\text{exp}\left(-\int_{0}^{r}a(s)ds\right)drdt' \\
 m_3 = &\text{exp}\left(\int_{0}^{t}a(r)dr\right)\int_{0}^{t}\text{exp}\left(\int_{0}^{t'}a(r)dr\right)
            \left[\int_{0}^{t'}\text{cos}(\omega r)\text{exp}\left(-\int_{0}^{r}a(s)ds\right)dr\right]^2 dt'.
\end{align}
Taking the inverse-Fourier transformation leads to
\begin{align}
 P_1(x,t)=-(m_1+m_2+c^2_1 m_3)\frac{\partial P_0}{\partial x}+c_1(2St_1+St_2)\frac{\partial^2 P_0}{\partial x^2}
                -\frac{1}{2}(s_1+s_2)\frac{\partial^3 P_0}{\partial x^3}.
\end{align}
Hence,
\begin{align}
 P(x,t)=P_0(x,t)-\sigma\left((m_1+m_2+c^2_1m_3)\frac{\partial P_0}{\partial x}-c_1(2St_1+St_2)\frac{\partial^2 P_0}{\partial x^2}
            +\frac{1}{2}(s_1+s_2)\frac{\partial^3 P_0}{\partial x^3}\right).
\end{align}

\subsection{\label{sec:periodic} Interpretation of the solution}
 The inclusion of long-term forcing leads to several new terms in each order. 
 First, in the O(1) order, long-term 
 forcing results in a deviation from the steady state solution. The deviation is represented 
 by $L$, which represents a combination of long-term forcing $\text{cos}(\omega t)$ and periodic stability $a(t)$.
 Under the assumption $\omega << 1/T$, where $T$ is the period of $a(t)$, $L$ is approximated as
 \begin{align}
  L &\simeq c_1 \text{cos}(\omega t)\int_{0}^{t}\text{exp}\left(\int_{t'}^{t}a(r)dr\right)dt \nonumber \\
     &=c_1 \text{cos}(\omega t) \text{WL}(t).
 \end{align}
$\text{WL}(t)$ can be interpreted as a weighting factor 
 controlled by the instantaneous stability, $a(t)$.
 As per the approximate solution, the response to long-term forcing has the same frequency as that of the forcing and its 
 magnitude is weighted by the memory effect shaped by $a(t)$.
  In the $O(1)$ order, the role of long-term forcing is to move the mean slowly following the phase of the forcing, whereas
the standard deviation remains same as in the case without long-term forcing. 

In the next order $O(\sigma)$, 
long-term forcing begins interacting with the nonlinearity of the system and the multiplicative structure of 
the noise, which changes the standard deviation of the stochastic solution. Moreover, a more complicated combination
of long-term forcing and the stability, $a(t)$, tends to update the mean of the solution. We look into these parts 
in detail.

First, we need to recollect that in the case without long-term forcing there is no change in standard deviation in the order \cite{MW:2013}.
The mean and the skewness are modified owing to the nonlinearity and multiplicative noise. The inclusion of 
long-term forcing makes it possible to change the standard deviation.
As a simple 
observation, we can see that $St_1$ is related to the interaction of long-term forcing with the nonlinearity and $St_2$
with the multiplicative noise structure. For a better understanding, we can use different forms of $St_1$ and $St_2$,
\begin{align}
St_1 &\simeq c_1\text{cos}(\omega t)\int_{0}^{t}\text{exp}\left(2\int_{t'}^{t}adr\right)\left[b(t')\right]\left[\int_{0}^{t'}\text{exp}
           \left(\int_{r}^{t'}ads\right)dr\right]  
         \times \left[\int_{0}^{t'}N^2(r)\text{exp}\left(2\int_{r}^{t'}ads\right)dr\right]dt' \nonumber \\
         &= c_1\text{cos}(\omega t) \text{WSt}_1(t) \nonumber \\
St_2 &\simeq c_1 \text{cos}(\omega t)\int_{0}^{t}\text{exp}\left(2\int_{t'}^{t}adr\right)\left[N(t')g(t')
          \right] 
         \times \left[\int_{0}^{t'}\text{exp}\left(\int_{r}^{t'}ads\right)dr\right]dt' \nonumber \\
        &=c_1\text{cos}(\omega t) \text{WSt}_2(t). 
\end{align}
Here, we assume that $\omega << 1/T$, such that $c_1\text{cos}(\omega t)$ could be out of the integral form
and  $\text{WSt}_1(t)$ and  $\text{WSt}_2(t)$ are the weightings controlled by a combination of the stability, $a(t)$, the nonlinearity, $b(t)$, 
and the noise structure, $N(t)g(t)$. For both terms, $\text{WSt}_1(t)$ and $\text{WSt}_2(t)$,
there is a common structure, which is 
\begin{align}
 \int_{0}^{t} \text{exp}\left(2\int_{t'}^{t}a(r)dr\right)\text{F}(t')dt',
\end{align}
where F is an arbitrary function. We would like to denote $\int_{0}^{t}\text{exp}\left(2\int_{t'}^{t}adr\right)\left[\cdot \right]dt' $ 
as the memory kernel.  The role of this kernel is to accumulate all past quantities till the present. The weighting is determined
based on the magnitude of the memory, $\exp(2\int_{t'}^{t}a(r)dr)$. Hence, the difference between $St_1$ and $St_2$ is what 
is accumulated. For $St_1$, 
\begin{align}
 F(t') = b(t') \times \int_{0}^{t'}\text{exp}\left(\int_{r}^{t'}ads\right)dr  \times \int_{0}^{t'}N^2(r)\text{exp}\left(2\int_{r}^{t'}ads\right)dr,
\end{align}
where $b(t')$ is the instantaneous nonlinearity, $\int_{0}^{t'}\text{exp}\left(\int_{r}^{t'}ads\right)dr$ the memory deposit 
 and $\int_{0}^{t'}N^2(r)\text{exp}\left(2\int_{r}^{t'}ads\right)dr$ the accumulation of the noise intensity. These three terms are 
 combined at time $t'$ and then accumulated again until the present, $t$. Similarly, $St_2$ is the result of the 
 accumulation of two terms, $N(t')g(t')$ and the memory deposit. It is worth noting that $St_1$ and $St_2$ can decrease
 or increase the standard deviation depending on the signs of $b(t)$ and $N(t)g(t)$ combined with the phase of the long-term 
 forcing.
 
 
 The final contribution of long-term forcing in the $O(\sigma)$ order is the deviation of the mean. In the previous order, we see 
 that the main contribution of long-term forcing is to move the mean slowly with its phase, while the magnitude is controlled 
 by the stability and its induced memory effect. In this order, a similar contribution exists, which is expressed by
 \begin{align}
  m_3 &\simeq c_1 \text{cos}(\omega t) \int_{0}^{t} \text{exp}\left(\int_{t'}^{t}adr\right)\left[\int_{0}^{t'}
           \text{exp}\left(\int_{r}^{t'}ads\right)dr\right]^2 dt' \nonumber \\
          &=c_1 \text{cos}(\omega t) \text{Wm}_3(t).
 \end{align}
 The weighting, $\text{Wm}_3(t)$, is also controlled by the stability and its memory effect. The difference between the current and the previous order is that the memory 
 effects are twice combined inside the integral. We also observed this structure in $St_1$ and $St_2$. We can expect that as the order
 increases, the memory effects tend to be more complicatedly interlaced inside the integral. 
 

 \section{\label{sec:seaice} Stochastic sea ice model with long-term forcing}
 In the previous research \cite{MW:2013}, a generalised periodic non-autonomous stochastic model is applied to the Arctic sea ice thickness model.
 Arctic sea ice serves as a good example in the field of climate science for demonstrating distinct seasonal characteristics in terms of monthly stability $a(t)$. Hence, 
 it expresses the difference between an autonomous stochastic model and a periodic non-autonomous one with reasonable accuracy. In this research, it is also desirable 
 to use the Arctic sea ice model as an example to emphasise the role of seasonality with long-term forcing. Further, the application 
 of the previous general theory can also be considered in the study of climate science. 
 
 \subsection{Analysis of the terms related to long-term forcing}
 A detailed model description can be found in \cite{EW09, MW:2013}. The sea ice model is based on the 
 energy heat flux balance between two boundaries in sea ice. On the boundary between sea ice and atmosphere, 
 radiative, sensible and latent heat fluxes are balanced by conductive heat flux in sea ice. Similarly, there is a balance between 
 oceanic turbulent heat flux and conductive heat flux on the boundary between ocean and sea ice. This overall balance is represented 
 by a single periodic non-autonomous ODE evolving the thickness of sea ice with seasonally-varying external fluxes. 
 This energy flux balance induces a seasonal cycle of sea ice thickness.
 To observe the effect of global warming on sea ice, 
 the additional heat flux, $\Delta F_0$, is included as a control parameter that ranges from $0.0W/m^2$ to $30.0W/m^2$.
 
 Short-time variabilities can be realised by stochastic noise forcing. The inclusion of 
 stochastic noise into the energy flux balance leads to a stochastic sea ice model.  The equation of the model is
 \begin{align}
  \frac{dE}{dt}=f(E,t)+\sigma b(E,t)\xi,
 \end{align} 
 with
 \begin{align}
  f(E,t)=F_D-F_T(t)T(t,E)+F_B+\mu R(-E)
 \end{align}
 where $F_D = [1-\alpha(E)]F_S(t)-F_0(t)+\Delta F_0$. The periodic function $F_S(t)$ represents short-wave radiative flux evolving seasonally. 
 The other term, $F_0(t)-F_T(t)T(t,E)$, contains longwave radiative flux, turbulent sensible and latent heat fluxes and meridional atmospheric 
 heat flux from low-latitudes. Short-wave radiative flux is controlled by the albedo, $\alpha(E)$, which is a function of the sea ice energy, $E$. 
 Oceanic turbulent heat flux is represented by the constant, $F_B$; further, 
 sea ice export out of the Arctic region, caused mainly by atmospheric large circulation, is also considered in the last term. 
 The ramp function, $R$, is used to initiate this term only when sea ice exists ( $E<0$).
 These time-functions are constructed based on monthly-averaged observation data.
 The last term on the right side, $\sigma b(E,t) \xi$, is stochastic forcing implying the effect of weather-related processes.
 Based on observation data, it is realistic to assume that $\sigma$ is quite small compared to the overall seasonal cycle of the main variable, $E$.
 $E$ represents the latent heat stored in the sea ice when it is negative and it is the thermal energy stored in the ocean mixed layer when $E$ is 
 positive. Hence, $f(E,t)$ is the energy flux sum over sea ice if $E$ is negative and when $E$ is positive, $f(E,t)$ is the energy flux over ocean mixed layer.
 
 \begin{figure}[htb]
\centering
\subfigure [] {
   \includegraphics[scale =0.33] {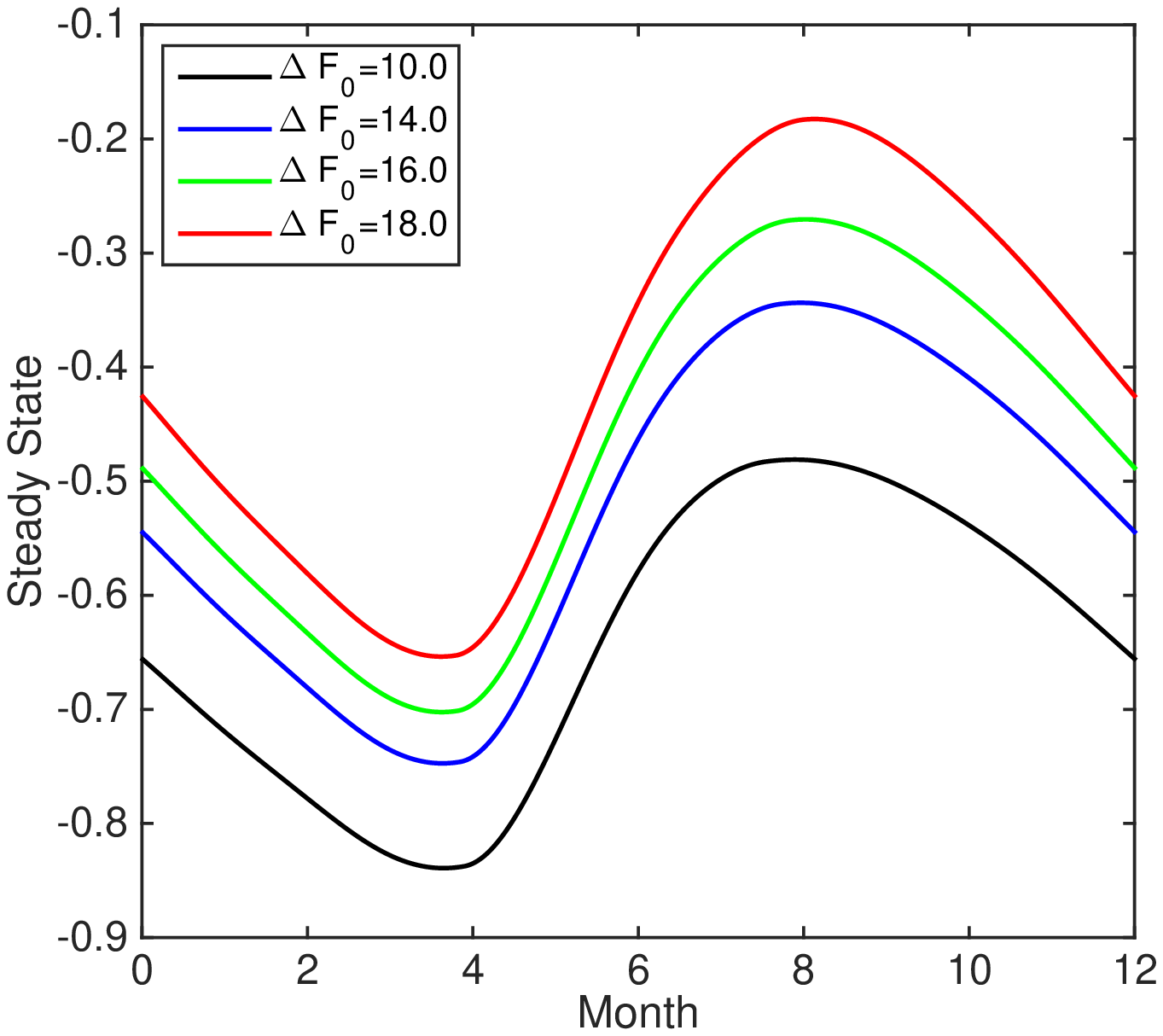}}
\subfigure[]{
   \includegraphics[scale =0.33] {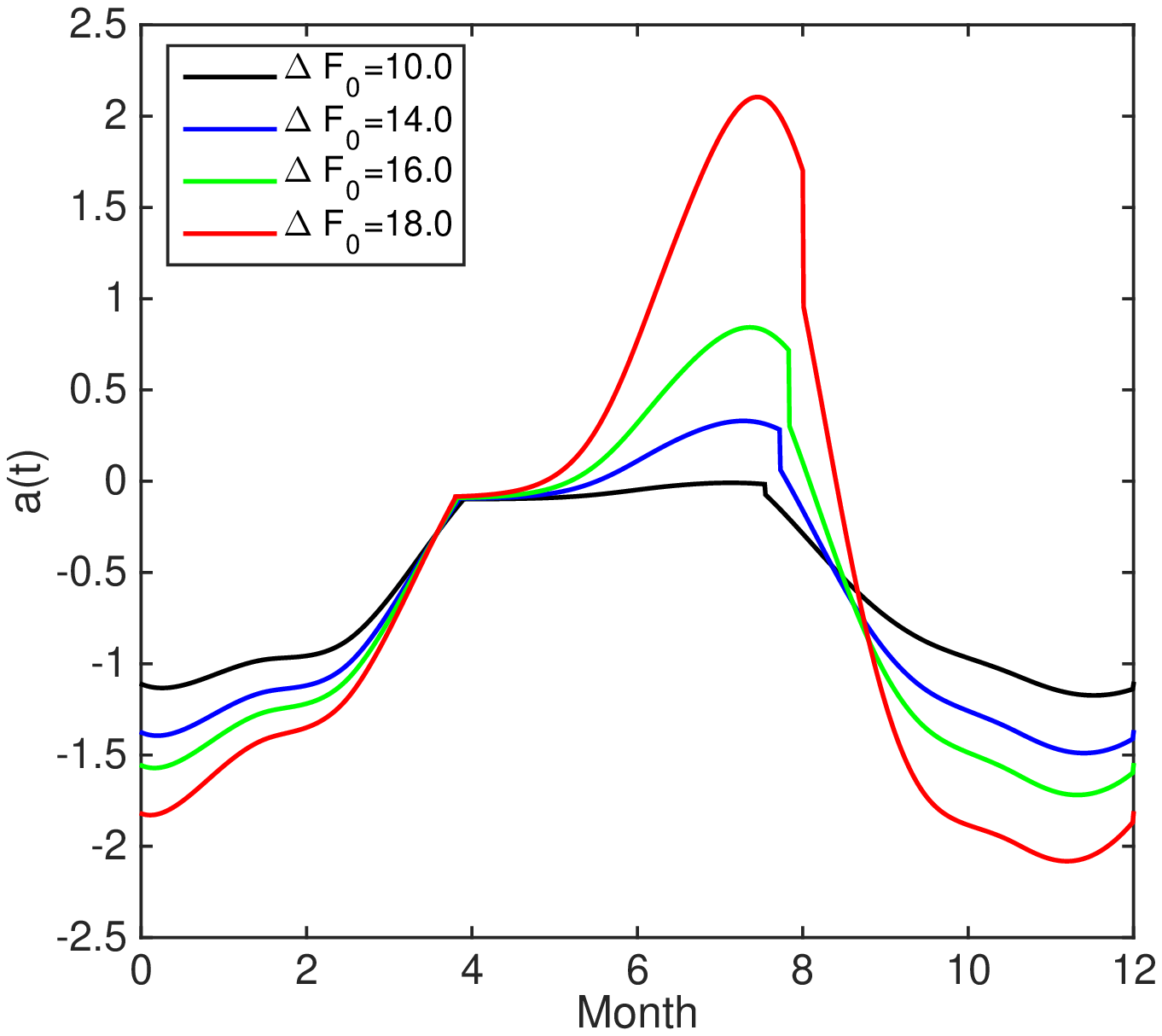}}
\subfigure[]{
    \includegraphics[scale=0.33] {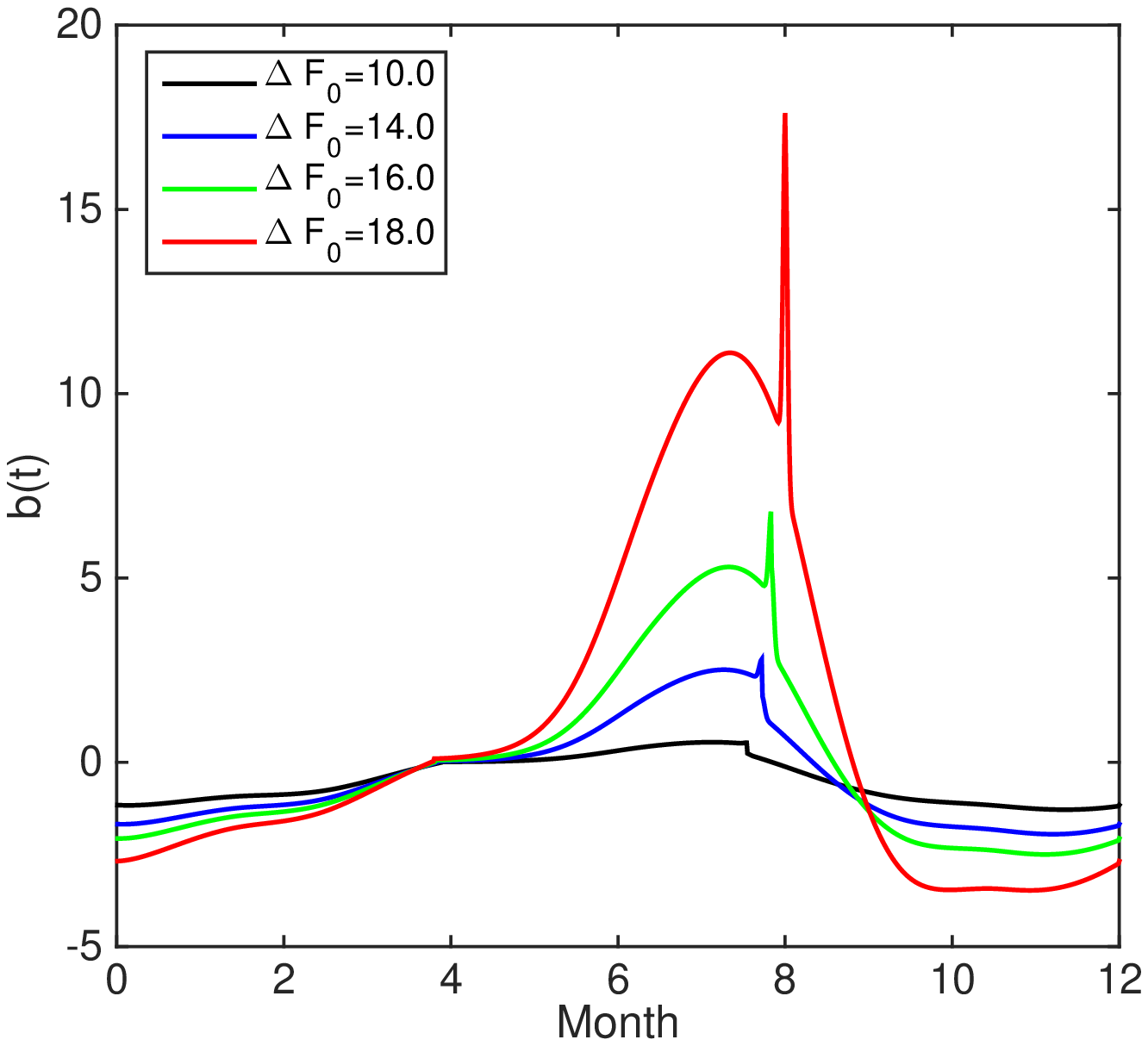}}
\caption{The characteristics of the steady state solutions of the sea ice model. The steady state solutions (a), the seasonal 
Sensitivity, $a(t)$ (b), and the nonlinearity, $b(t)$ (c) with different additional heat flux, $\Delta F_0$, are shown. }
\label{fig:seaice}
\end{figure}

Examples of the steady state solutions, the seasonal sensitivity, $a(t)$, and the nonlinearity, $b(t)$, are shown in figure \ref{fig:seaice}. 
The equation is non-dimensionalised, owing to the latent heat stored in 3 m sea ice ($L_i h_0$), where $L_i$ is the heat fusion of sea ice 
and $h_0 = 3m$. In the figure, we can see that the thickness of sea ice reflected in the steady state solutions decreases 
as the additional heat, $\Delta F_0$, increases. Furthermore, as $\Delta F_0$ increases, the absolute magnitude of $a(t)$ also 
decreases, which indicates the weakening of the stability of sea ice; then, $a(t)$ becomes even positive during summer when 
$\Delta F_0$ is 18.0. The nonlinearity, $b(t)$, shows more seasonality when $\Delta F_0$ increases, which is related to 
the asymmetric response of the sea ice albedo feedback during summer.
   
 Based on the 
 assumption that we already know the periodic steady state solution of the deterministic part, $E_S(t)$, we can consider a Taylor expansion
 around the steady state, $E_S$, to observe the variability around the solution due to small stochastic forcing. Hence, we obtain
 \begin{align}
  &a(t)=\frac{\partial f(E,t)}{\partial E}|_{E=E_S} \nonumber \\
  &b(t)=\frac{1}{2}\frac{\partial^2 f(E,t) }{\partial E^2}|_{E=E_S} \nonumber \\
  &N(t)=b(E_S,t) \nonumber \\
  &g(t)=\frac{\partial b(E,t)}{\partial E}|_{E=E_S}, 
 \end{align}
 which leads to
 \begin{align}
 \frac{d\eta}{dt}=a(t)\eta+b(t)\eta^2+\sigma (N(t)+g(t)\eta)\xi
 \end{align}
 after introducing $\eta=E-E_S$. We can now introduce long-term forcing, which has the same order of magnitude as that of the noise. Our equation becomes
 \begin{align}
 \frac{d\eta}{dt}=a(t)\eta+b(t)\eta^2+\sigma (N(t)+g(t)\eta)\xi+\sigma c_1 \text{cos}(2 \pi t).
 \end{align}
 
 The periodic time-functions $a(t)$, $b(t)$, $N(t)$, and $g(t)$ are different depending on control parameter $\Delta F_0$. As $\Delta F_0$ increases, which 
 could be interpreted as the effect of on-going global warming, the seasonality of $a(t)$ is intensified and the magnitude of $b(t)$, interpreted as the 
 asymmetry of the energy flux balance of Arctic sea ice increases. Detailed information regarding the change of these periodic functions can be found 
 in the previous research \cite{MW:2013}.
 
\begin{figure}[!h]
\centering
\subfigure [] {
   \includegraphics[scale =0.40] {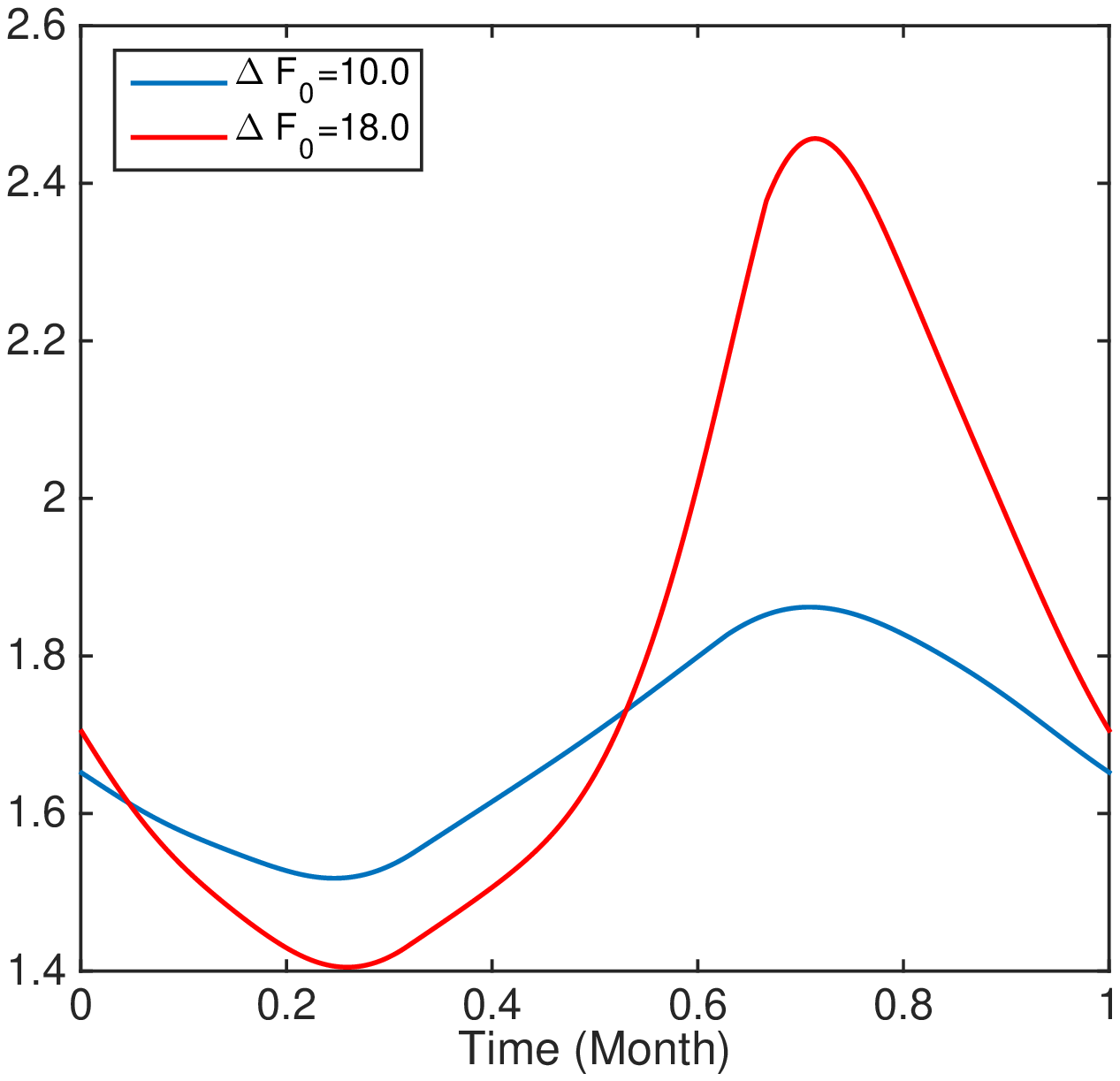}}
\subfigure[]{
   \includegraphics[scale =0.40] {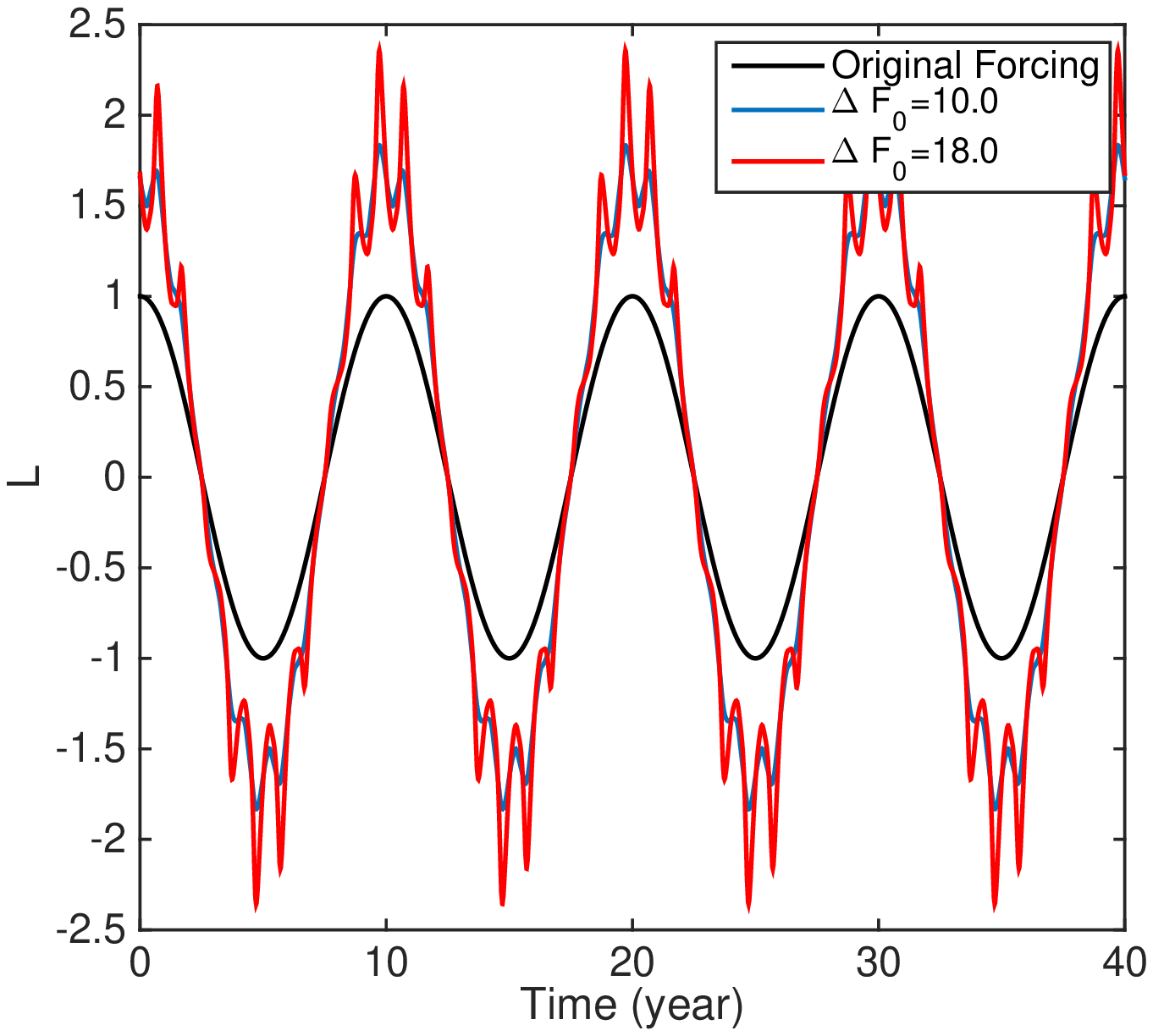}}
\caption{The deviation of the mean owing to long-term periodic forcing ($L$). Figure \ref{fig:seaice}(a) shows the seasonal weighting, $\text{WL}(t)$, 
with two different $\Delta F_0$s. 
The overall
long-term change caused by forcing is shown in figure \ref{fig:seaice}(b). Here, the seasonal period is $1.0$ and the period 
of the long-term forcing is $10.0$. In figure \ref{fig:seaice}(b), the black line indicates the original forcing. Compared with figure \ref{fig:seaice}(a),
we can see that the response to forcing increases owing to the weighting controlled by the memory effect.}
\label{fig:L}
\end{figure}

 We investigated the role of long-term forcing based on the theory. In the first order, long-term forcing moves
 the mean with its phase. The magnitude of the deviation of the mean owing to long-term forcing is controlled by stability $a(t)$.
Figure \ref{fig:L} shows the evolution of the mean owing to long-term periodic forcing magnified by the seasonal contribution
 of $a(t)$. In figure \ref{fig:L} (a), we observe the seasonal change of weighting $\text{WL}(t)$
 with two different $\Delta F_0$s. In particular, for a higher $\Delta F_0$, the seasonal variability is large owing to an intensified sea ice albedo during summer. 
 In the overall evolution 
 of $L$ (shown in b), the seasonal variability is more magnified during extreme phases. As $a(t)$ has larger seasonal variability ($\Delta F_0=18.0$),
 the response to long-term forcing during extreme phases shows larger fluctuations. This is owing to a combination of the memory effect 
 and the magnitude of the long-term forcing.  
 
 \begin{figure}[!h]
\centering
\subfigure [] {
   \includegraphics[scale =0.40] {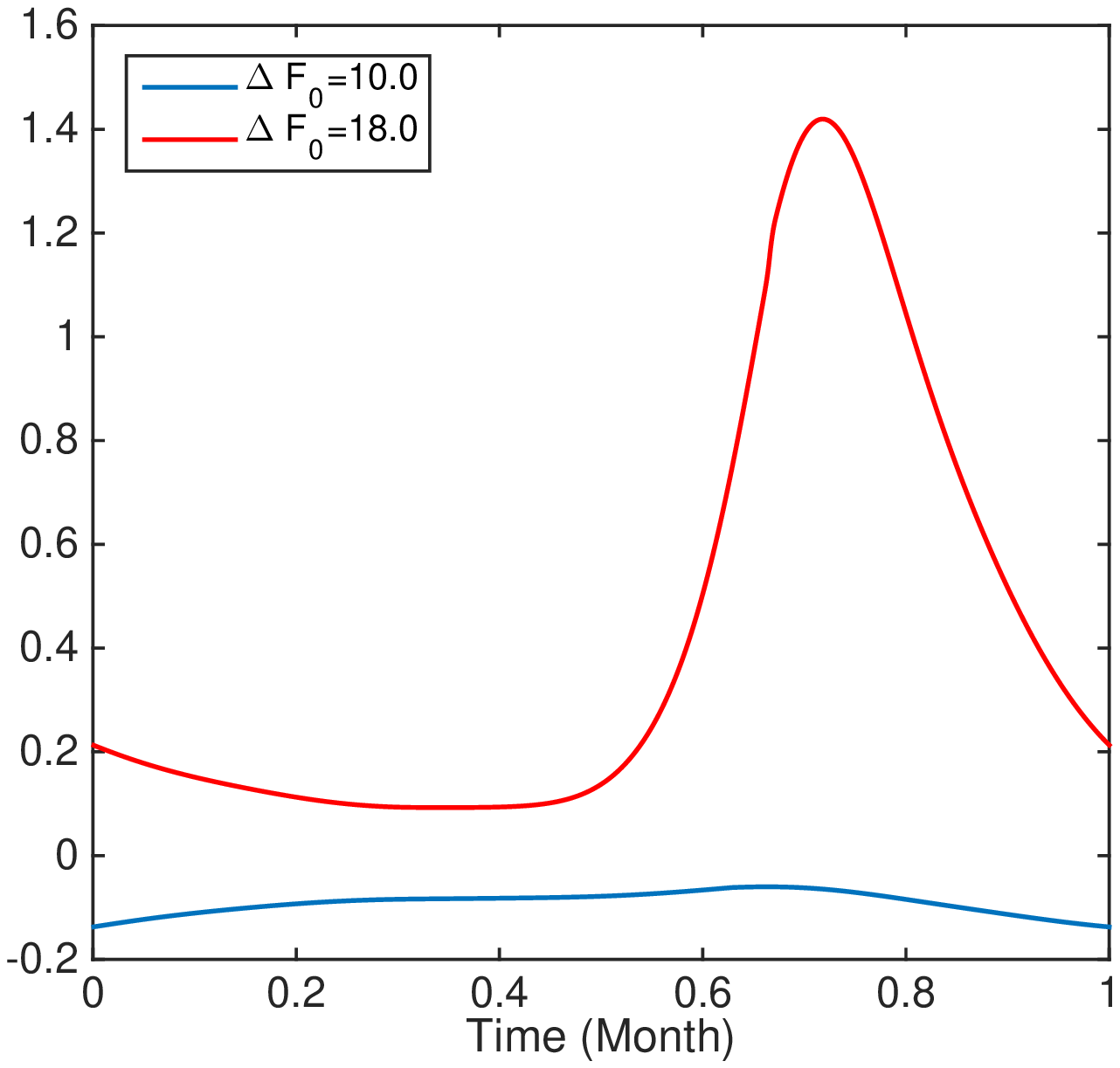}}
\subfigure[]{
   \includegraphics[scale =0.40] {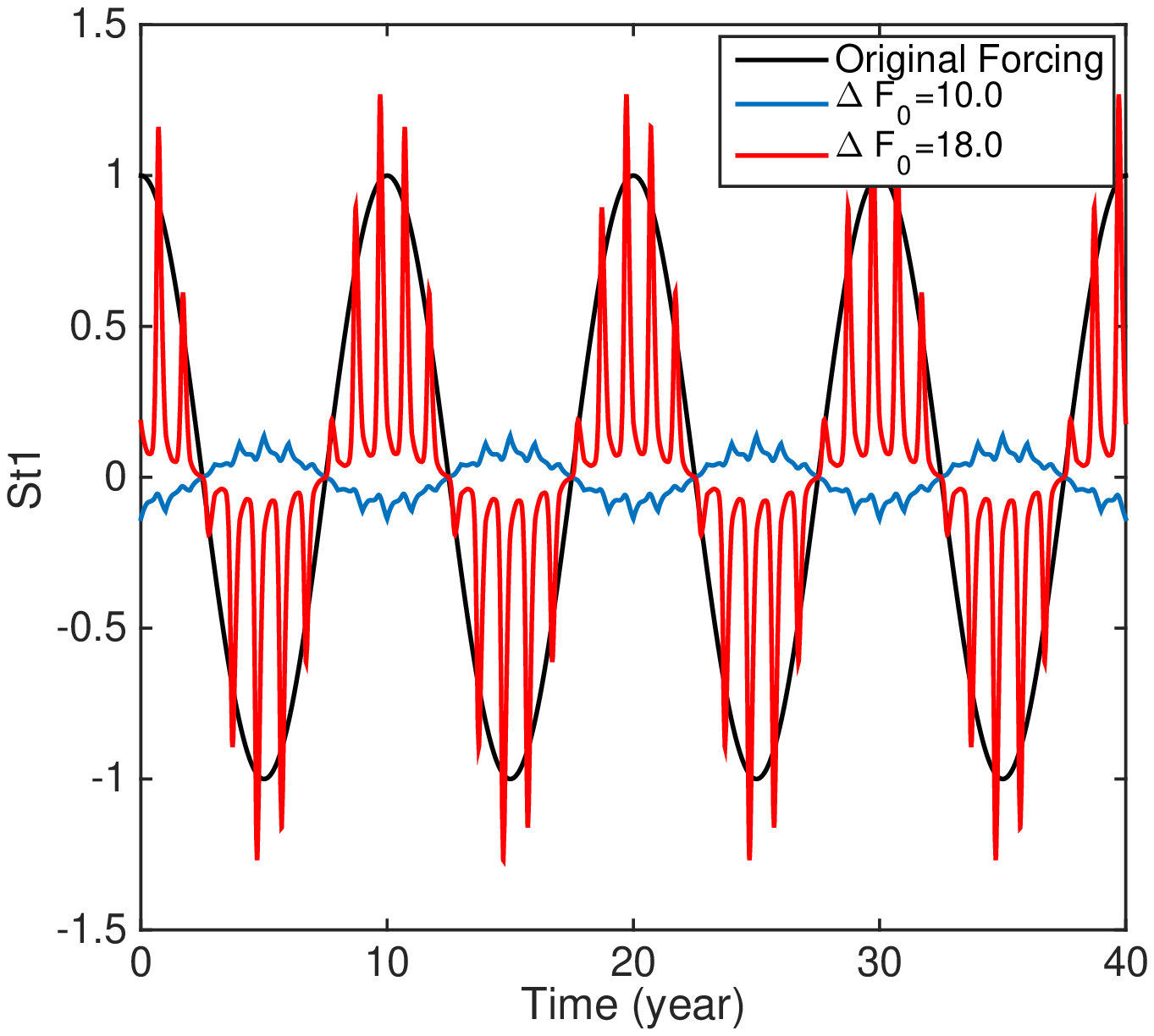}}
 \subfigure [] {
   \includegraphics[scale =0.40] {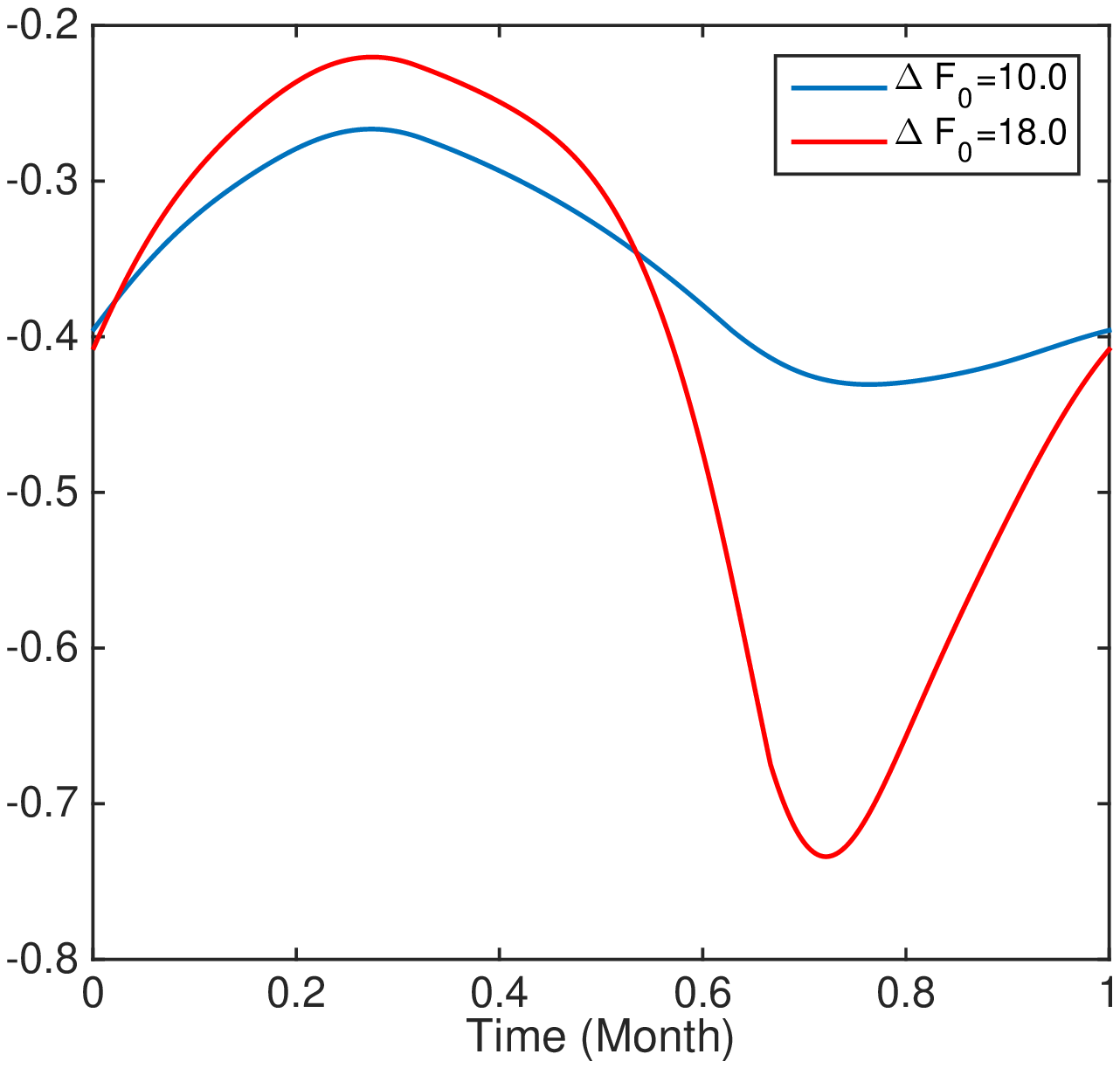}}
\subfigure[]{
   \includegraphics[scale =0.40] {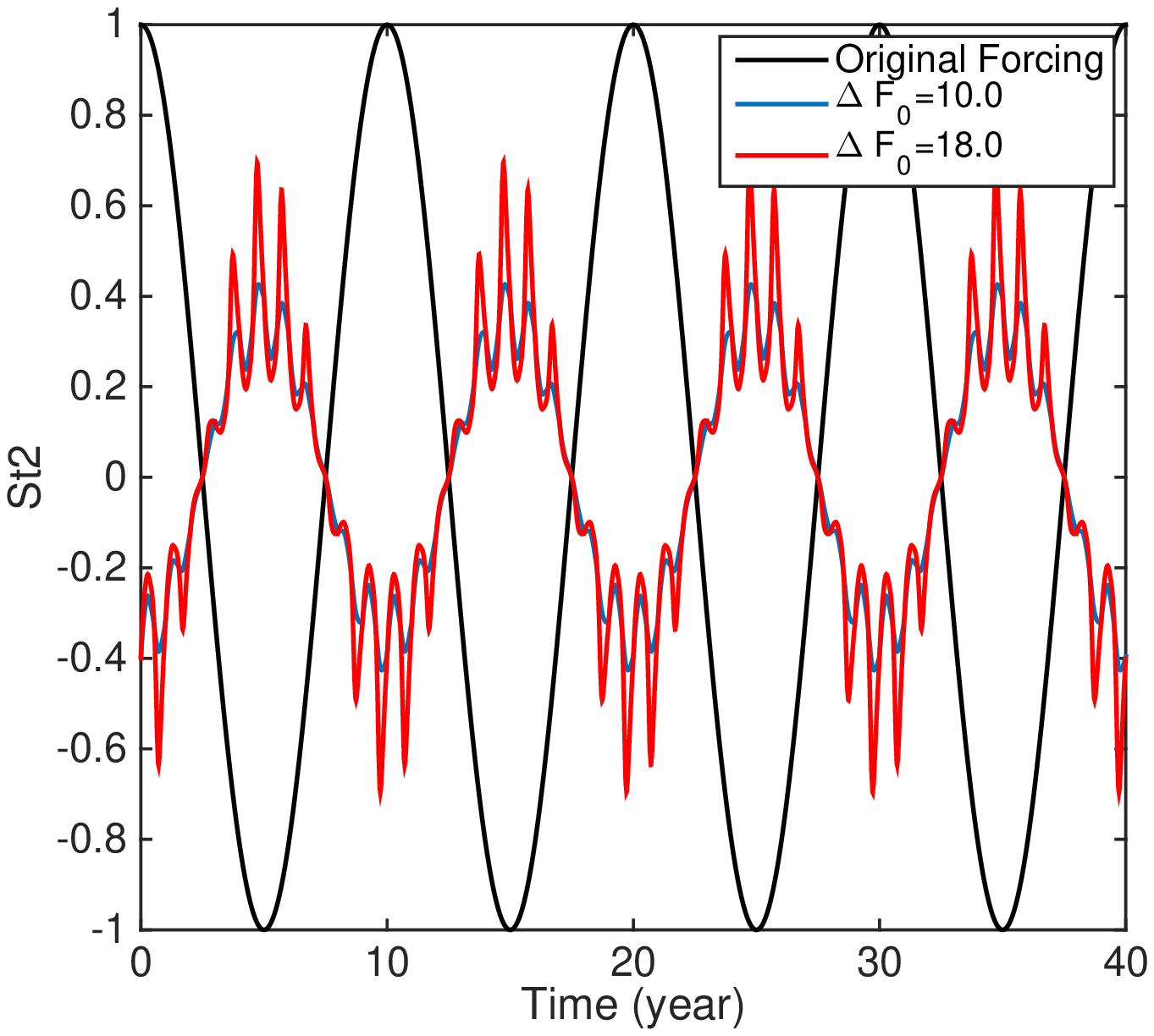}}
  
\caption{The standard deviation changes owing to long-term periodic forcing represented by $St_1$ (b) and $St_2$ (d). The weighting 
factor controlled by the memory effect combined with the noise and the nonlinearity $\text{WSt}_1(t)$ is shown in (a) for two $\Delta F_0$s. 
The other weighting factor generated by the memory effect associated with multiplicative noise $\text{WSt}_2(t)$ is in (c).
The seasonal terms, $\text{WSt}_1(t)$ and $\text{WSt}_2(t)$,
are multiplied to the original long-term forcing shown by the black curves in (b) and (d).   }
\label{fig:st}
\end{figure}

The change of the standard deviation owing to long-term forcing is also shown in figure \ref{fig:st}. The interaction of long-term 
forcing with the nonlinearity, $b(t)$, and with the multiplicative noise are represented in (b) and (d), respectively. The multipliers, $\text{WSt}_1(t)$
and $\text{WSt}_2(t)$, induced
from $a(t)$ and $b(t)$ or $N(t)g(t)$ to the long-term forcing are shown in (a) and (c), respectively. Depending on the sign and magnitude
of $b(t)$ or $N(t)g(t)$, the original signal represented as the black curve in (b) and (d) are magnified positively or negatively. For $N(t)g(t)$,
we choose $-1$ to represent the variability of the Arctic sea ice export out of the Arctic basin from the Fram Strait \cite{MW:2013}. The common features are also shown 
in these cases. The response to long-term forcing becomes significantly  larger during extreme phases of forcing.

\begin{figure}[!h]
\centering
\subfigure [] {
   \includegraphics[scale =0.40] {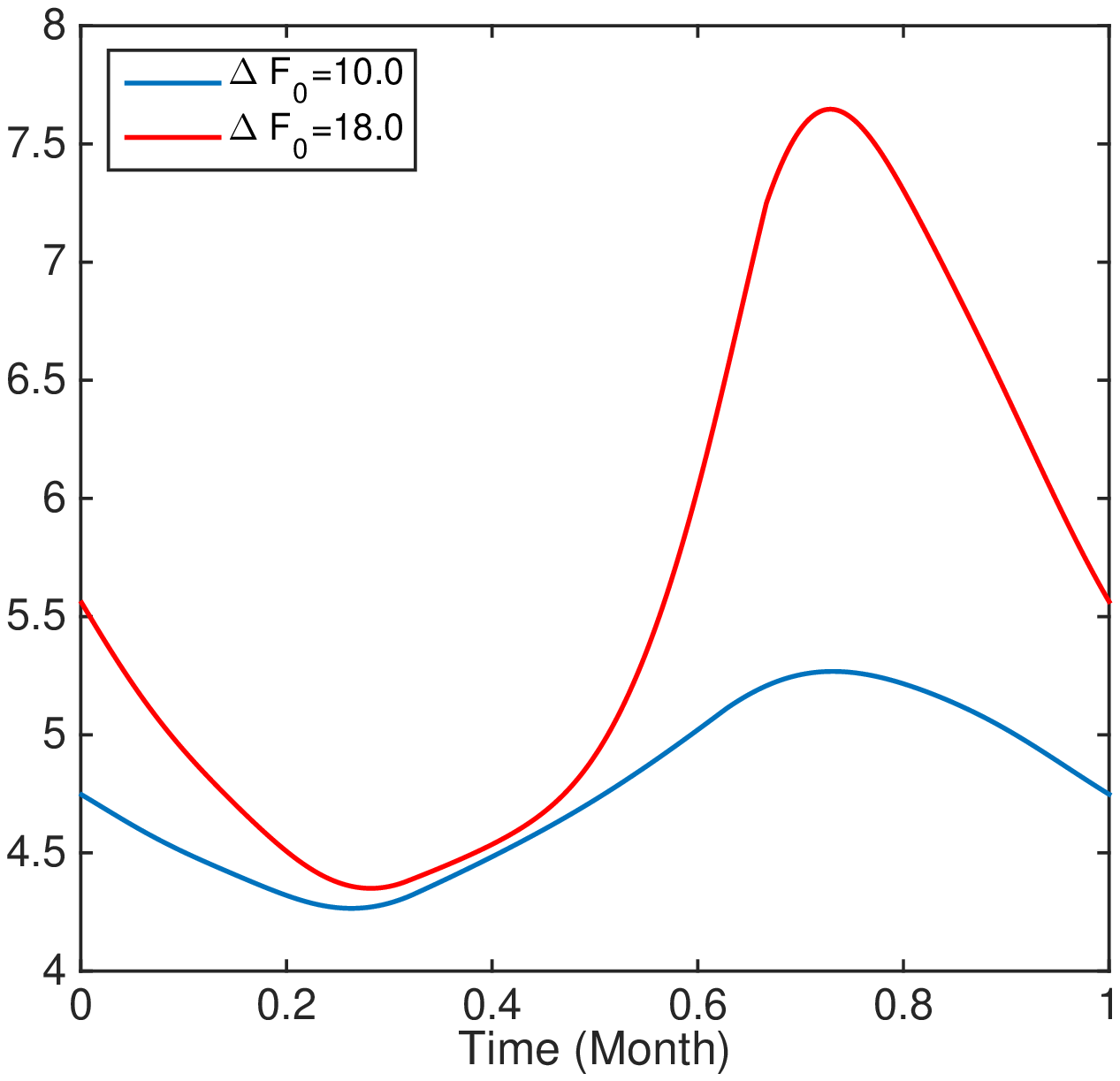}}
\subfigure[]{
   \includegraphics[scale =0.40] {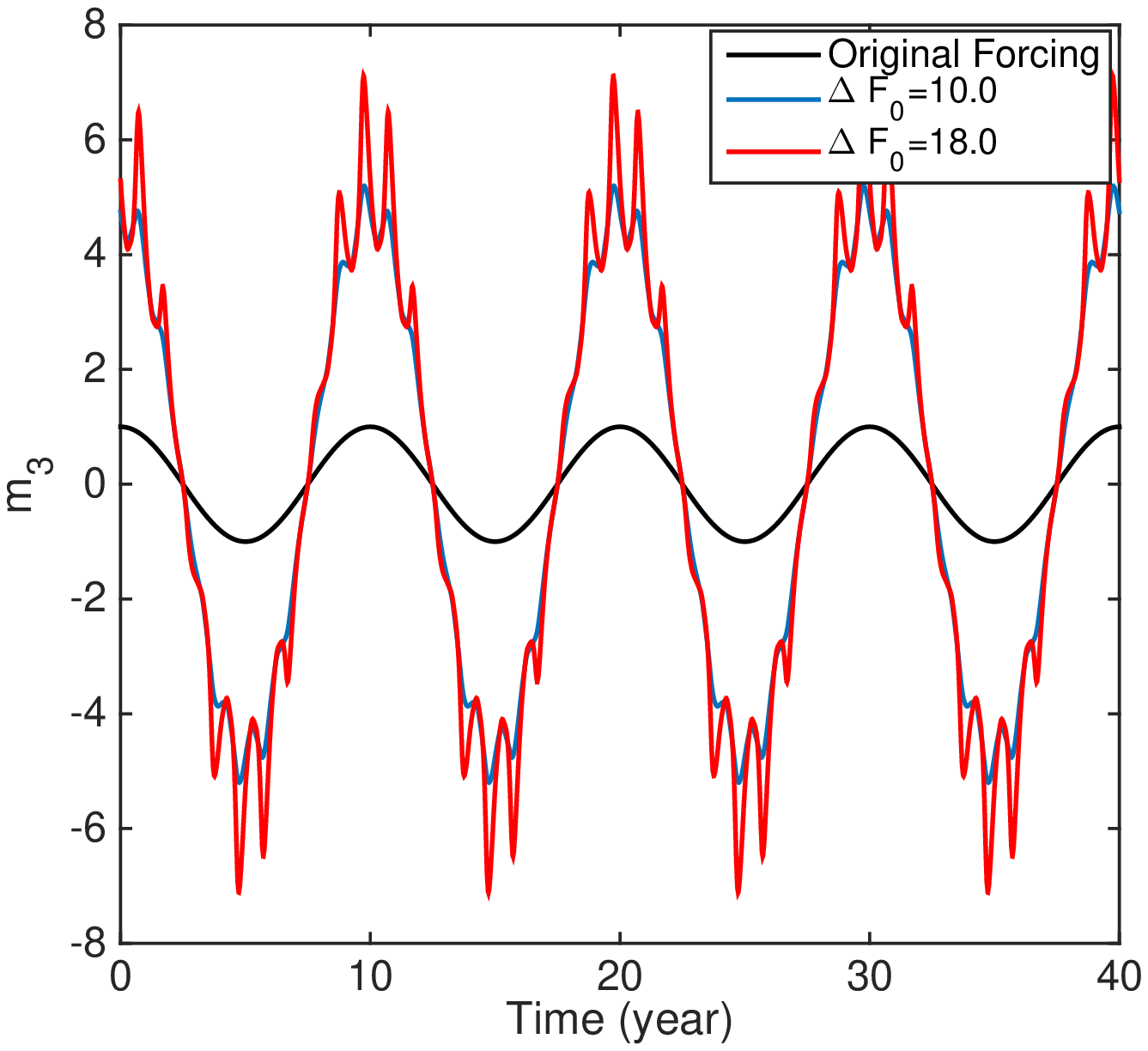}}
\caption{The deviation of the mean owing to long-term periodic forcing in the $O(\sigma)$ order is shown in (b). 
(a) shows the seasonal weighting, $\text{Wm}_3(t)$,
which is induced from $a(t)$ and then influences the magnitude of the effect of long-term forcing. }
\label{fig:m3}
\end{figure}

The last term in consideration is the deviation of the mean, $m_3$, in the $O(\sigma)$ order. Figure \ref{fig:m3} shows 
similar information as the previous figures. As $\Delta F_0$ increases, the seasonal variability becomes larger. 
The overall response to long-term forcing is maximised near extremes, similar to other quantities. 

From the stochastic Arctic sea ice model, we investigated the role of long-term forcing reflected in several statistical moments terms in each order.
In the stochastic Arctic sea ice model, the solutions became less stable and had larger seasonality as the additional heat flux $\Delta F_0$ 
increased. The seasonality of the stability, nonlinearity, and multiplicative noise characteristics interacts with long-term forcing, 
in which case the response of the seasonality is maximised near the extremes of the long-term forcing. 

\subsection{The role of the long-term forcing in time-series}

The inclusion of long-term forcing into the seasonal stochastic model influences a change in seasonal statistics through the interaction of 
long-term forcing with the stability, $a(t)$, the nonlinearity, $b(t)$, and the multiplicative noise structure, $N(t)g(t)$. The above theory yields approximate
estimations of several statistical moments, standard deviation, mean, and skewness. 
We need to observe how the change in statistical moments by the inclusion of long-term forcing is projected in time-series.

\begin{figure}[!hbtp]
\centering
\includegraphics[angle=0,scale=0.38,trim= 0mm 0mm 0mm 0mm, clip]{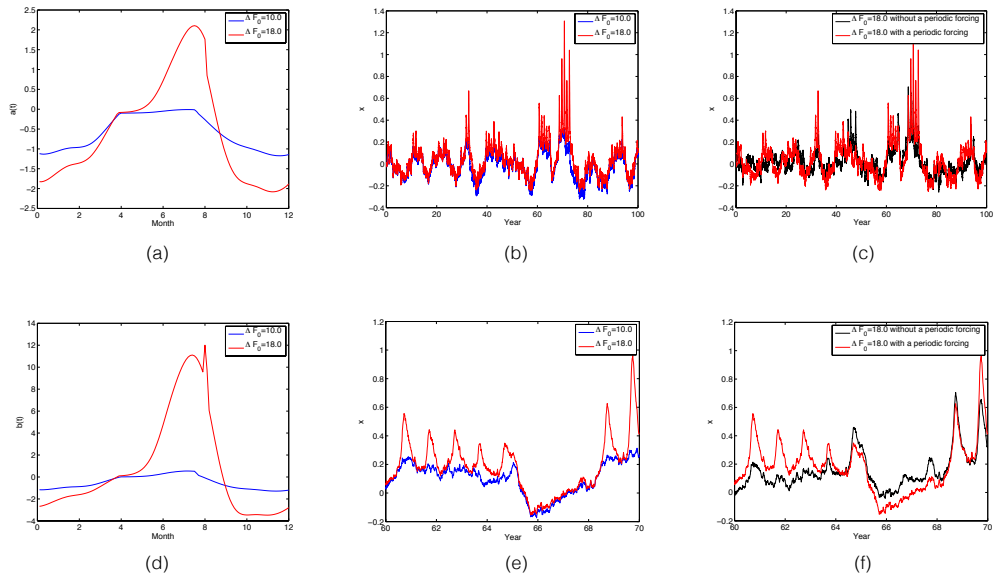}
\caption{The comparison of $a(t)$ and $b(t)$ for the two $\Delta F_0$s, 10.0 and 18.0, is shown in (a) and (d). The comparison 
of the time-series generated from $a(t)$s and $b(t)$s for the two $\Delta F_0$s is shown in (b) and (e). The blue curve is for 10.0 and the 
red one for 18.0. In (e), the same time-series is shown in a narrower time domain for observing the difference more closely. 
The two cases with long-term forcing (red) and without long-term 
forcing (black) are also compared in (c) and (f). A narrower time-domain view is also shown in (f).}
\label{fig:ex_nonlinear}
\end{figure}

To observe the effect of long-term forcing interacting with the seasonality in time-series, we need to compare stochastic realisations from several stochastic 
models.
The test stochastic models are  
\begin{align}
 &\frac{dx}{dt}=a_1(t)x+b_1(t)x^2+\sigma \xi + c_1 \text{cos}(\omega t) \nonumber \\
 &\frac{dx}{dt}=a_2(t)x+b_2(t)x^2+\sigma \xi + c_1 \text{cos}(\omega t) \nonumber \\
 &\frac{dx}{dt}=a_2(t)x+b_2(t)x^2+\sigma \xi,
\end{align}
where $a_1(t)$ and $b_1(t)$ are constructed from the Arctic sea ice model for $\Delta F_0=10.0$ and $a_2(t)$ and $b_2(t)$ are constructed for
$\Delta F_0=18.0$. 
The two cases have very similar $\int_{0}^{T}a(r)dr$, which is around 
$-0.59$. The only difference is the degree of seasonality. The third one, the case without long-term forcing, 
is also considered for observing the effect of long-term forcing by comparison with 
the second model. For the comparison, all of the models implement the same random numbers at each time step so that the differences among the three models
are exclusively from their different structures, and not from any random effects.
$a(t)$s and $b(t)$s for the two $\Delta F_0$s are shown in figure \ref{fig:ex_nonlinear} (a) and (d).

In figure \ref{fig:ex_nonlinear} (b) and (e), we can observe a comparison for stochastic realisations between two different $\Delta F_0$s.  
The difference between the two cases lies on the degree of seasonality in the stability, $a(t)$, and the nonlinearity, $b(t)$. In particular, 
the magnitude of $b(t)$ is quite different for the two cases. According to the theory, long-term forcing is intertwined with $b(t)$ in the memory kernel for changing the variance. The effect of the combination between long-term forcing and $b(t)$
 upon the time-series is realised by the sharp peaks 
in the time-series (the red curve in (b)). In the narrow time domain, shown in (e), it is clearly seen that the increase or decrease of fluctuations is associated
with the phase of long-term forcing. In this comparison, it is found that the interaction between the seasonality and 
long-term forcing results in the occurrence of large fluctuations phased with long-term forcing.

The other comparisons shown in (c) and (f) also suggest the role of long-term forcing more clearly. The black curve is a case without long-term 
forcing and the red one with long-term forcing. Because the magnitude of long-term forcing is not larger than noise forcing, the shift of the mean
is not well detected owing to the existence of noise forcing. Instead, the contribution of long-term forcing is represented by
the intensified sharp signals following the phase of long-term forcing (red curve in (c) and (f)).  
The other distinct characteristics is that the intensified signals shown in sharp peaks are concentrated near the end of summer in the seasonal time domain.
Owing to the memory effect, the extreme signals are shown to be concentrated when the seasonal variance is maximised. This is normally near the end of 
summer for the Arctic sea ice.
 
\begin{figure}[!hbtp]
\centering
\includegraphics[angle=0,scale=0.38,trim= 0mm 0mm 0mm 0mm, clip]{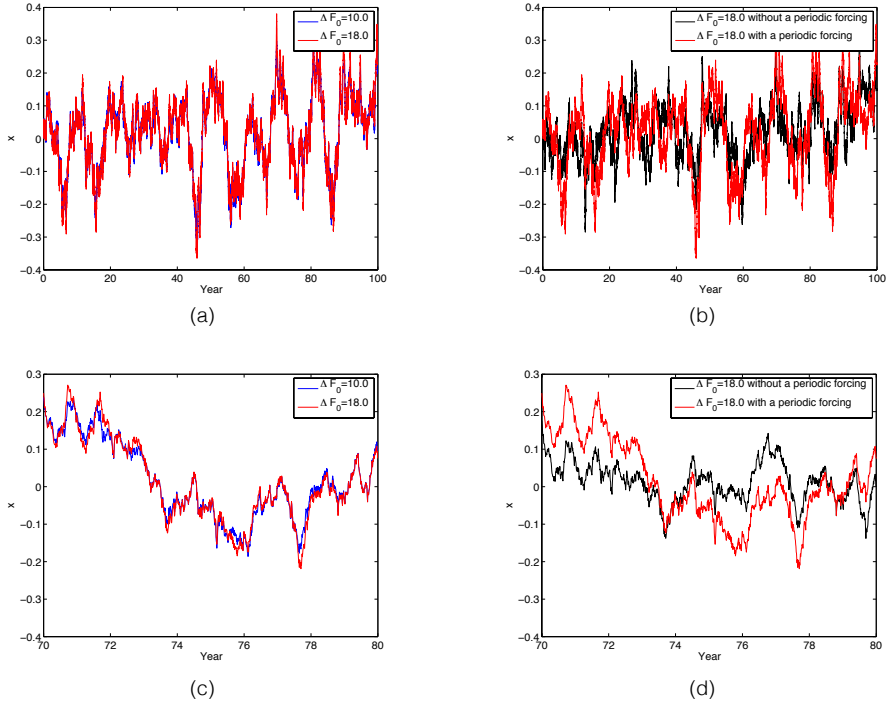}
\caption{The comparison of the time-series generated from two different $\Delta F_0$s is represented in (a) and (c). A narrower time domain
view is shown in (c). Another comparison with long-term forcing and without long-term forcing is in (b) and (d), respectively. In (d), the time-series can 
be observed in a narrower time domain.}
\label{fig:ex_multi}
\end{figure}

We also have to observe the situation wherein long-term forcing interacts with multiplicative noise structure. 
According to the theory, long-term forcing also interacts with multiplicative noise to shift the variance. We can rely on 
a similar comparison to observe the role of long-term forcing with multiplicative noise. 

The three models can be suggested as
\begin{align}
  &\frac{dx}{dt}=a_1(t)x+\sigma (1-x) \xi + c_1 \text{cos}(\omega t) \nonumber \\
  &\frac{dx}{dt}= a_2(t)x +\sigma (1-x) \xi + c_1 \text{cos}(\omega t) \nonumber \\
   &\frac{dx}{dt}= a_2(t)x +\sigma (1-x) \xi, 
\end{align}
 where $a_1(t)$ and $a_2(t)$ are used again and we use that $N(t)=1$ and $g(t)=-1$. 
 The results are shown in figure \ref{fig:ex_multi}. The comparison between the two different $\Delta F_0$s is shown in (a) and (c) along with
a magnified view in a narrow time domain in (c), to observe the difference more closely. The blue line is for $\Delta F_0=10.0$ and the red one for $\Delta F_0=18.0$.
 However, the difference between the two cases is not significant compared to the case with nonlinearity, $b(t)$, with long-term forcing. 
 Another comparison between a case with long-term forcing (red) and the one without long-term forcing (black) is shown in (b) and (d), respectively.
 These two cases use the same random numbers for time advancing, and thus, short-time fluctuations appear almost the same. The distinct difference lies in
 the increase or decrease of the fluctuations depending on the phase of long-term forcing. However, the difference between the two cases 
 is not distinctive compared with the previous case, which shows the interaction of long-term forcing with $b(t)$.
 
In this section, the results imply that the interaction of long-term forcing with nonlinearity or multiplicative noise in the stochastic Arctic sea ice model
results in large unusual fluctuations in the seasonal time domain 
when the memory effect is maximised. Owing to the small magnitude of long-term forcing compared with the size of 
background noise, it may be hard to detect the existence of long-term forcing using a simple spectrum analysis. The existence of long-term forcing 
seems to be realised by unusually large
fluctuations at a specific time of a year. 
 
\section{Conclusions}  

Seasonality is a fundamental feature that needs to be considered, primarily when one observes any physical variables in our climate. Seasonality is not an internal
characteristic of a climate system but rather an externally-induced one. Hence, periodic non-autonomous characteristics are inevitable in 
all of daily and monthly data spanning several decades. 
In this research, we investigate how seasonality can interact with 
long-term forcing based on a simple periodic non-autonomous stochastic model. 

Previously, a regular perturbation method is applied to a periodic non-autonomous stochastic ordinary differential equation with small noise 
based on the rules of stochastic calculus. On the other hand, this research uses the equivalent Fokker-Planck equation, for which 
we rely on a singular perturbation method, Fourier transformation, and the characteristics method. We prove that this different approach 
also yields the same approximate solutions as the previous one. The advantage of this method 
is its ability to directly calculate the probability density function, which enables us to avoid complicated calculations with stochastic calculus in high orders.

We include a simple long-term forcing into the given periodic non-autonomous stochastic equation. We apply the Fokker-Planck equation formalism 
to find out stochastic solutions. Simple long-term forcing modifies the original solutions. First, the stochastic mean changes slowly 
with the same phase as that of the long-term forcing. This effect is not necessarily unique in a periodic non-autonomous model but a common feature
with an autonomous one. Moreover, if long-term forcing is not significantly larger than background noise, the mean shift is not observed as a 
distinct feature. 

The uniqueness of the periodic non-autonomous stochastic ODE with long-term forcing is in the $O(\sigma)$ order.
The standard deviations change owing to long-term forcing combined with nonlinearity and multiplicative noise. Without long-term 
forcing, only the first two odd statistical moments are affected by the nonlinearity and multiplicative noise structure in the order. 
However, the inclusion of long-term forcing leads to a change in the standard deviation by the interaction of the long-term forcing 
with the nonlinearity and the multiplicative noise.
The mean shift also exists in this order, but the expression is more complicated in its interaction with the memory effect. The next issue is determining how 
a change in standard deviation and the mean in this order is shown in the time-series generated by the periodic non-autonomous stochastic ODE
with long-term forcing.

For this, the stochastic Arctic sea ice model is used again to see the effect of long-term forcing in stochastic realisations. The weakened stability and 
intensified nonlinearity during summer
combined with the extreme phases of long-term forcing generates sharp peaks in time series. 
The findings indicate that the unusual signals or events in a particular season could be due to the interaction of the intensified nonlinearity or the multiplicative noise
with the phase of long-term forcing. Even though it is almost impossible to observe the slow change of the mean in time-series, the occurrence of 
the unusual peaks during a specific time of a year may be used to determine the existence of long-term forcing statistically.
Moreover, we may explore the possibility that the knowledge of low-frequency leads to the prediction of unusual events for a specific year.

Owing to the complexity of the various scale interactions in our climate, it is difficult to say that the simple model approach represents the reality in a full range, but this result
suggests important qualitative aspects of the seasonality influenced by a myriad of slowly varying physical processes. With different seasonal 
stability and nonlinearity and weather-like short-time processes, long-term forcing could contribute to not only long-term variability but also seasonal variability.
In particular, long-term forcing may be related to the occurrence of extreme events represented as sharp peaks.
In the future, it would be beneficial to extend this research with more complicated and realistic considerations.


\begin{thebibliography}{9}

\bibitem{MW:2011} Moon, W. and J. S. Wettlaufer, 2011: A low-order theory of Arctic sea ice stability.
\emph{Europhys. Lett.}, \textbf{96}, 39001 (doi: 10.1209/0295-5075/96/39001)

\bibitem{EW09} Eisenman, I. and J. S. Wettlaufer, 2009: Nonlinear threshold behavior during the loss of Arctic sea ice.
\emph{Proc. Natl. Acad. Sci. USA}, \textbf{106}, 28-32. (doi: 10.1073/pnas.0806887106)

\bibitem{MW:2013} Moon, W. and J. S. Wettlaufer, 2013: A stochastic perturbation theory for non-autonomous systems.
\emph{J. Math. Phys.}, \textbf{54(12)}, 123303. 

\bibitem{MW:2012} Moon, W. and J .S. Wettlaufer, 2012: On the existence of stable seasonally varying Arctic sea ice in simple models.
\emph{J. Geophys. Res.-Oceans}, \textbf{117(C7)}.

\bibitem{Branstator1992} Branstator, G., 1992: The maintenance of low-frequency atomspheric anomalies.
\emph{J. Atmos. Sci.}, \textbf{49(20)}, 1924-1946.

\bibitem{Hurrell1995} Hurrell, J. W., 1995: Decadal trends in the North Atlantic Oscillation: regional temperatures and precipitation.
\emph{Science}, \textbf{269(5224)}, 676-679.


\bibitem{Agarwal2011} Agarwal, S., W. Moon and J. S. Wettlaufer, 2011: Decadal to seasonal variability of Arctic sea ice albedo.
\emph{Geophys. Res. Lett.}, \textbf{38(20)}

\bibitem{Agarwal2012} Agarwal, S., W. Moon and J. S. Wettlaufer, 2012: Trends, noise and re-entrant long-term persistence in Arctic sea ice.
\emph{Proc. R. Soc. A.}, \textbf{468(2144)}, 2416-2432
  
\end{thebibliography}
\end{document}